\documentclass[aps,prc,twocolumn,nofootinbib,amsmath,superscriptaddress,floatfix,10pt]{revtex4-1}
\usepackage{dcolumn}
\usepackage{graphicx}
\usepackage{mathptmx}
\usepackage{epstopdf}  
\usepackage[pdftex,colorlinks,linkcolor = blue,urlcolor  = blue,citecolor = blue]{hyperref}
\usepackage{hyperref}
\usepackage{latexsym}

\begin{document}

\title{Uncertainties in radiative neutron-capture rates relevant to the $A\sim 80$ $r$-process peak}

\author{I.~K.~B.~Kullmann}
\email{i.k.b.kullmann@fys.uio.no}
\affiliation{Department of Physics, University of Oslo, N-0316 Oslo, Norway}
\author{E.~W.~Hafli}
\affiliation{Department of Physics, University of Oslo, N-0316 Oslo, Norway}
\author{A.~C.~Larsen}
\email{a.c.larsen@fys.uio.no}
\affiliation{Department of Physics, University of Oslo, N-0316 Oslo, Norway}
\author{E.~Lima}
\affiliation{Department of Physics, University of Oslo, N-0316 Oslo, Norway}

\date{\today}

\begin{abstract}
The rapid neutron-capture process ($r$-process) has for the first time been confirmed to take place in a neutron-star merger event. A detailed understanding of the rapid neutron-capture process is one of the holy grails in nuclear astrophysics.
In this work we investigate one aspect of the $r$-process modelling:
uncertainties in radiative neutron-capture cross sections and astrophysical 
reaction rates for isotopes of the 
elements Fe, Co, Ni, Cu, Zn, Ga, Ge, As, and Se. 
In particular, we study deviations from standard libraries used for astrophysics, and the influence
of a very-low $\gamma$-energy enhancement in the average, reduced $\gamma$-decay probability on the 
($n,\gamma$) rates. We find that the intrinsic uncertainties are in some cases extremely large, and that the low-energy enhancement, if present in neutron-rich nuclei, may increase the neutron-capture reaction rate significantly. 
\end{abstract}


\maketitle

\section{introduction}
\label{sec:int}
One of the \textit{"Eleven Science Questions for the New Century"}~\cite{questions2003} concerns the
formation of elements from iron to uranium in stellar environments. Although the main processes
responsible for the heavy-element creation were outlined already in 1957 by 
Burbidge \textit{et al.}~\cite{burbidge1957} and Cameron~\cite{cameron1957}, 
there are at present many open questions regarding the heavy-element nucleosynthesis.
This is particularly true for the \textit{rapid neutron capture (r-) process},
which is known to produce about half of the observed abundances from Fe to U. The $r$-process must take
place in extreme, astrophysical environments with a very high neutron flux;
in order to reach uranium ($A=238$) from iron-group seed nuclei ($A\sim 50-100$), 
the $r$-process site needs to provide about 100 neutrons per seed nucleus~\cite{meyer1994}.

For many decades, the astrophysical site for the $r$-process was not uniquely identified~\cite{arnould2007,thielemann2011}; both supernovae~\cite{hillebrandt1976,woosley1994} and compact-object mergers~\cite{lattimer1974,lattimer1977,Freiburghaus1999} were suggested sources for the observed $r$-process material. 
Initially, simplified simulations of core-collapse supernovae indicated that they could provide conditions favorable for an $r$-process~(\textit{e.g.} Ref.~\cite{hillebrandt1976}), which created a lot of excitement.
However, more realistic simulations allowing for e.g. spherical asymmetry and with updated
neutrino physics were unsuccessful both in terms of actually making a core-collapse supernova explode, as well as creating a sufficiently neutron-rich, high-entropy environment~(see, e.g., Ref.~\cite{janka2012}
and references therein). Further, the so-called neutrino-driven wind stemming from the nascent neutron star that remains after the core-collapse supernova has been a very popular site to explore for $r$-process nucleosynthesis (e.g. Refs.~\cite{woosley1994} and references therein). Recent simulations including 
a detailed neutrino-transport treatment~\cite{martinez-pinedo2012} led only to
a slightly neutron-rich environment at the early stages after the bounce (first$\sim 3$ seconds),
while at later times the ejecta turn out to be proton rich. Of course, as there are, so far, no
supernova simulations taking all known physics ingredients properly into account~\cite{goriely2016},
supernovae can still not be excluded as possible producers of $r$-process material.

On August 17, 2017, a huge breakthrough was achieved as the $r$-process was observed live for the first time.  
the Advanced LIGO and Advanced Virgo gravitational-wave detectors made the first discovery of a neutron-star binary merger named GW170817~\cite{LIGO2017}. 
The gravitational-wave observation was immediately followed up by measuring the electromagnetic radiation of GW170817 over a broad range of frequencies. A short $\gamma$-ray burst from GW170817 with duration of $\approx2$s was detected by the \textit{Fermi} Gamma-Ray Burst Monitor~\cite{Fermi2017} and the INTEGRAL telescope~\cite{INTEGRAL2017}. 
Furthermore, measurements for several weeks in the ultraviolet, (near-)optical and infrared wavelengths~\cite{Coulter2017,Valenti2017,Pian2017,Drout2017} revealed that the "afterglow" ("also called "macronova" or "kilonova") of GW170817 is fully compatible with the expected light curve powered by radioactive decay of heavy isotopes~\cite{Kasen2017} produced in the $r$-process. According to Kasen \textit{et al.}~\cite{Kasen2017}, the observations fit well with models assuming two components of $r$-process ejecta: one for the light masses ($A <140$) and one for heavier masses ($A\geq 140$).  
 
On the modelling side, significant process has also been made recently: detailed simulations
of neutron star mergers and neutron star$-$black hole mergers provide a convincing abundance pattern
very similar to the solar-system $r$-process abundances for $A>130$ nuclei~\cite{goriely2013,just15}.
That said, a detailed and realistic treatment of all physics ingredients, such as neutrino transport and magnetic fields,
is still lacking in the neutron-star merger models~\cite{goriely2016}.

From a nuclear-physics point of view, the required amount of nuclear data needed for an $r$-process
simulation is overwhelming. 
For a detailed $r$-process nucleosynthesis network simulation, input nuclear data include masses, 
$\beta$-decay rates, radiative neutron-capture rates, and fission properties.
It has proven very difficult to identify key nuclei or mass regions of special importance
for an $r$-process calculation, as this may vary considerably for different astrophysical conditions. 
However, regardless of the astrophysical site,
it is clear that the $r$-process involves many 
unstable nuclei with large neutron-to-proton ratios and very short lifetimes.  
For sophisticated $r$-process simulations, $\approx$ 5,000 nuclei and $\approx$ 50,000 reaction rates 
must be included in the reaction network. 
  
The majority of these neutron-rich nuclei are currently out of reach
experimentally, although a significant mass-range extension opens up when new facilities
such as FRIB~\cite{FRIB}, FAIR~\cite{FAIR} and HIE-ISOLDE~\cite{ISOLDE} will be fully operational. Even then,
performing direct measurements of $(n,\gamma)$ cross sections will be virtually impossible.
Hence, one must rely on theory to provide the necessary cross sections for the nucleosynthesis
simulations. Providing stringent tests and experimental constraints on the theoretical calculations
would be of utmost importance to ensure reliability and robustness of the calculated cross sections.

A sensitivity study by Surman \textit{et al.}~\cite{surman2014} indicate that some 
$(n,\gamma)$ reaction rates for $A \sim 80$ nuclei have a significant impact on the final abundances
in this mass region for a large set of different astrophysical conditions, varying the entropy and electron fraction,
and varying the $(n,\gamma)$ reaction rates by a factor of 100. 
As pointed out in Ref.~\cite{surman2014}, reducing the uncertainties in the nuclear input is critical
for an accurate prediction of the abundance patterns. This conclusion is further underlined in 
a recent review of Mumpower \textit{et al.}~\cite{mumpower2016}, where the huge uncertainties in 
neutron-capture rates are emphasized $-$ different predictions can  easily vary by orders of magnitude 
for isotopes away from stability. 

To calculate $(n,\gamma)$ cross sections, three nuclear ingredients are of key
importance: the level density (NLD), the $\gamma$ strength function ($\gamma$SF), and the neutron optical-model
potential (n-OMP)~\cite{arnould2007}. 
For exotic nuclei far from stability, different mass models will also lead to significantly different results. 
In this work, we investigate the impact on these quantities
for moderately neutron-rich nuclei in the $A\sim 70-90$ mass region using the nuclear-reaction
code TALYS-1.8~\cite{TALYS_16,koning12}. In particular, we focus on the spread in predicted 
$(n,\gamma)$ cross sections and reaction rates using the implemented NLD, $\gamma$SF and n-OMP models
in TALYS. For nuclei where no measured masses are available, we vary the mass models as well. 
We also investigate the impact of including a low-energy enhancement in the $\gamma$SF — the so-called
\textit{upbend} seen in experimental data of stable nuclei this mass region
(e.g. Refs.~\cite{voinov2004,guttormsen2005,larsen2007,wiedeking2012,renstrom2016}). It has been shown 
previously~\cite{larsen_goriely_2010} that this enhancement may lead to a significant increase in the 
$(n,\gamma)$ reaction rates for very neutron-rich Fe, Mo, and Cd isotopes. 

In Ref.~\cite{larsen_goriely_2010},
it was assumed that the low-energy enhancement could be described as part of the $E1$ $\gamma$SF component,
and that it persisted throughout the isotopic chains all the way to the neutron drip line. The first 
experimental proof
that the low-energy upbend is indeed present in neutron-rich nuclei is in the experimental $\gamma$SF of
$^{70}$Ni~\cite{liddick2016,larsen2017}, showing a clear
increase in strength at $E_\gamma < 3$ MeV. These data have been applied to calculate the $^{69}$Ni($n,\gamma$)$^{70}$Ni 
reaction rate~\cite{liddick2016}, a first experimental constraint on this rate. However, although experimentally
shown to be dominated by dipole transitions~\cite{larsen2013}, it is at present not clear whether the 
upbend is due to magnetic or electric transitions, and so the underlying mechanism is not yet understood.
Theoretically, an $E1$ enhancement is found in calculations within the quasi-particle random phase approximation,
where thermal single-particle excitations are causing extra strength for low-energy $E1$ transitions~\cite{litvinova2013}. 
On the other hand, 
several shell-model calculations~\cite{schwengner2013,brown2014,schwengner2017,Sieja2017} demonstrate a strong increase in the $M1$ component 
of the $\gamma$SF at low transition energies. It remains to be experimentally verified whether the upbend is
indeed made up of $E1$ transitions, $M1$ transitions, or a mix of both. In this work, we assume it is 
of $M1$ type, and we study what impact it may have on the ($n,\gamma$) rates for neutron-rich
Fe, Co, Ni, Cu, Zn, Ga, Ge, As, and Se isotopes.

The paper is organized in the following way. In Sec.~\ref{sec:calcstable} we provide information about the 
applied inputs in TALYS and the calculations for nuclei near the valley of stability and 
compare the TALYS predictions with experimental data. In Sec.~\ref{sec:exotic}, we present our studies 
of the nuclei identified in Ref.~\cite{surman2014} to have maximum neutron-capture rate sensitivity measures, 
and compare with the recommended rates from the JINA REACLIB~\cite{JINA-REACLIB} 
and BRUSLIB~\cite{BRUSLIB} libraries. Finally, a summary and outlook are given in Sec.~\ref{sec:sum}.

\section{Cross-section calculations for stable and near-stable nuclei}
\label{sec:calcstable}
\begin{table*}[tb]
\begin{center}
\caption{Nuclei considered in this work for which experimental $(n,\gamma)$ cross-section data exist, 
and the TALYS keywords giving the lowest and highest predicted cross section. }
\begin{tabular}{lcll}
\hline
\hline
Reaction                        & Reference(s)                       & TALYS keywords, max & TALYS keywords, min \\
                                &                                    &                     &                     \\
\hline
$^{56}$Fe($n,\gamma$)$^{57}$Fe  & \cite{macklin1964,allen1976,shcherbakov1977} & \textit{ldmodel 6, strength 2, localomp n} & \textit{ldmodel 1, strength 1, jlmomp y} \\
$^{57}$Fe($n,\gamma$)$^{58}$Fe  & \cite{macklin1964,allen1977}                 & \textit{ldmodel 3, strength 2, localomp n} & \textit{ldmodel 1, strength 1, jlmomp y}\\
$^{58}$Fe($n,\gamma$)$^{59}$Fe  & \cite{trofimov1985,allen1980}                & \textit{ldmodel 3, strength 2, localomp n} & \textit{ldmodel 1, strength 1, jlmomp y}\\
$^{59}$Co($n,\gamma$)$^{60}$Co  & \cite{paulsen1967,spencer1976}               & \textit{ldmodel 6, strength 2, localomp n} & \textit{ldmodel 1, strength 7, jlmomp y}\\
$^{58}$Ni($n,\gamma$)$^{59}$Ni  & \cite{zugec2014,perey1993,guber2010}         & \textit{ldmodel 6, strength 2, localomp n} & \textit{ldmodel 3, strength 1, jlmomp y}\\
$^{60}$Ni($n,\gamma$)$^{61}$Ni  & \cite{stieglitz1971,perey1982}               & \textit{ldmodel 6, strength 2, localomp n} & \textit{ldmodel 1, strength 1, jlmomp y}\\
$^{61}$Ni($n,\gamma$)$^{62}$Ni  & \cite{tomyo2005}                             & \textit{ldmodel 2, strength 2, localomp n} & \textit{ldmodel 5, strength 1, jlmomp y}\\
$^{62}$Ni($n,\gamma$)$^{63}$Ni  & \cite{tomyo2005,alpizar2008}                 & \textit{ldmodel 2, strength 2, localomp n}& \textit{ldmodel 1, strength 7, jlmomp y}\\
$^{63}$Ni($n,\gamma$)$^{64}$Ni  & \cite{lederer2014,weigand2015}               & \textit{ldmodel 2, strength 2, localomp n} & \textit{ldmodel 4, strength 1, jlmomp y}\\
$^{64}$Ni($n,\gamma$)$^{65}$Ni  & \cite{grench1965,leipunskiy1958,booth1958}   & \textit{ldmodel 2, strength 2, localomp n} & \textit{ldmodel 1, strength 1, jlmomp y} \\
$^{63}$Cu($n,\gamma$)$^{64}$Cu  & \cite{blair1944,voignier1992,tolstikov1966,zaikin1968,kim2007,anand1979,booth1958,lyon1959}& \textit{ldmodel 2, strength 2, localomp n}& \textit{ldmodel 5, strength 1, jlmomp y} \\
$^{65}$Cu($n,\gamma$)$^{66}$Cu  & \cite{johnsrud1959,voignier1992,zaikin1968,leipunskiy1958,pasechnik1958,stavisskiy1961,tolstikov1964,lyon1959,macklin1957,peto1967,chaubey1966,colditz1968}  & \textit{ldmodel 3, strength 2, localomp n}& \textit{ldmodel 1, strength 1, jlmomp y} \\
$^{64}$Zn($n,\gamma$)$^{65}$Zn  & \cite{jinxiang1995,reifarth2012,schuman1970} & \textit{ldmodel 3, strength 2, localomp n} & \textit{ldmodel 1, strength 1, jlmomp y}\\
$^{66}$Zn($n,\gamma$)$^{67}$Zn  & \cite{garg1981}                              & \textit{ldmodel 3, strength 2, localomp n} & \textit{ldmodel 5, strength 1, jlmomp y}\\
$^{68}$Zn($n,\gamma$)$^{69}$Zn  & \cite{leipunskiy1958,booth1958,chaubey1966,colditz1968,garg1982,dovbenko1973,kononov1958}& \textit{ldmodel 3, strength 2, localomp n} & \textit{ldmodel 1, strength 1, jlmomp y}\\
$^{70}$Zn($n,\gamma$)$^{71}$Zn  & \cite{reifarth2012}                          & \textit{ldmodel 2, strength 2, localomp n} & \textit{ldmodel 1, strength 1, jlmomp y}\\
$^{69}$Ga($n,\gamma$)$^{70}$Ga  & \cite{dovbenko1969,zaikin1971,stavisskii1959,chaubey1966,kononov1958,pasechnik1958,peto1967}& \textit{ldmodel 3, strength 2, localomp n} & \textit{ldmodel 1, strength 1, jlmomp y}\\
$^{71}$Ga($n,\gamma$)$^{72}$Ga  & \cite{johnsrud1959,dovbenko1969,zaikin1971,stavisskii1959,liyang2012,anand1979,chaubey1966,lyon1959,macklin1957,peto1967,trofimov1987} & \textit{ldmodel 3, strength 2, localomp n} & \textit{ldmodel 6, strength 1, jlmomp y}\\
$^{74}$Ge($n,\gamma$)$^{75}$Ge  & \cite{tolstikov1967,pasechnik1958,anand1979,lyon1959,macklin1957,trofimov1987}  & \textit{ldmodel 3, strength 2, jlmomp y} & \textit{ldmodel 1, strength 1, localomp n}\\
$^{75}$As($n,\gamma$)$^{76}$As  & \cite{johnsrud1959,anand1979,booth1958,lyon1959,macklin1957,peto1967,colditz1968,beghian1949,chaturvedi1970,diksic1970,hasan1968,hummel1951,macklin1963,weston1960}  & \textit{ldmodel 3, strength 2, localomp n} & \textit{ldmodel 1, strength 1, jlmomp y}\\
$^{78}$Se($n,\gamma$)$^{79}$Se  & \cite{igashira2011}               & \textit{ldmodel 3, strength 2, localomp n} & \textit{ldmodel 5, strength 1, localomp n}\\
$^{80}$Se($n,\gamma$)$^{81}$Se  & \cite{tolstikov1967,igashira2011,sriramachandramurty1973,walter1984}  & \textit{ldmodel 4, strength 2, jlmomp y} & \textit{ldmodel 6, strength 1, localomp n}\\
$^{82}$Se($n,\gamma$)$^{83}$Se  & \cite{trofimov1987,igashira2011}                 & \textit{ldmodel 2, strength 2, localomp n} & \textit{ldmodel 6, strength 1, localomp n}\\
\hline
\hline
\end{tabular}
\label{tab:list_stable}
\end{center}
\end{table*}
\begin{figure*}[!hbt]
\begin{center}
\includegraphics[clip,width=1.3\columnwidth]{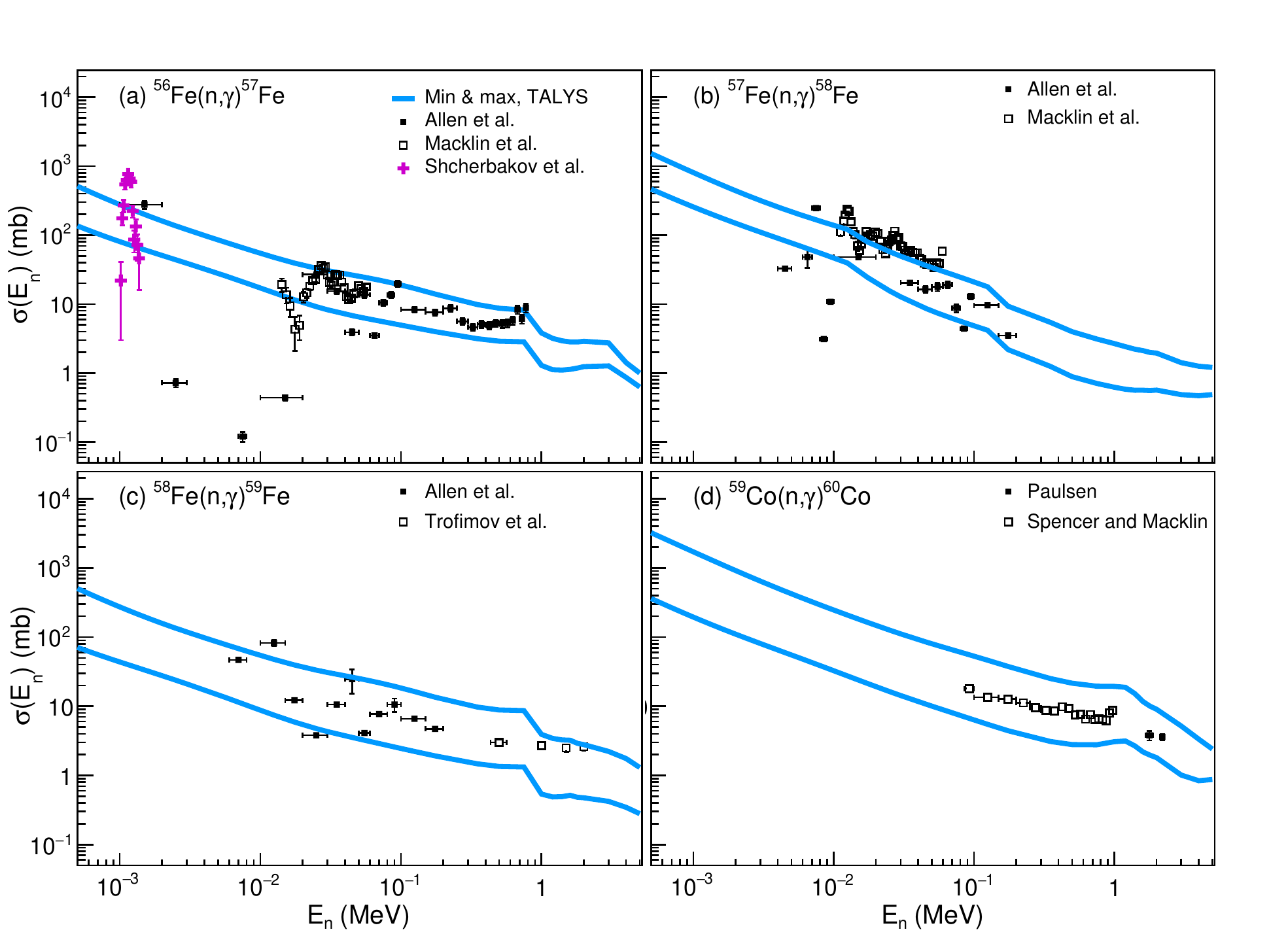}
\caption{(Color online) Comparison of measured (n,$\gamma$) cross sections with the minimum and maximum TALYS
predictions for (a) $^{56}$Fe(n,$\gamma$), data from Refs.~\cite{macklin1964,allen1976,shcherbakov1977}; 
(b) $^{57}$Fe(n,$\gamma$), data from Refs.~\cite{macklin1964,allen1977}; (c) $^{58}$Fe(n,$\gamma$), 
data from Refs.~\cite{trofimov1985,allen1980}, and (d) $^{59}$Co(n,$\gamma$), data from Refs.~\cite{trofimov1985,allen1980}.}
\label{fig:xsec_Fe_Co}
\end{center}
\end{figure*}
\begin{figure*}[!htb]
\begin{center}
\includegraphics[clip,width=1.6\columnwidth]{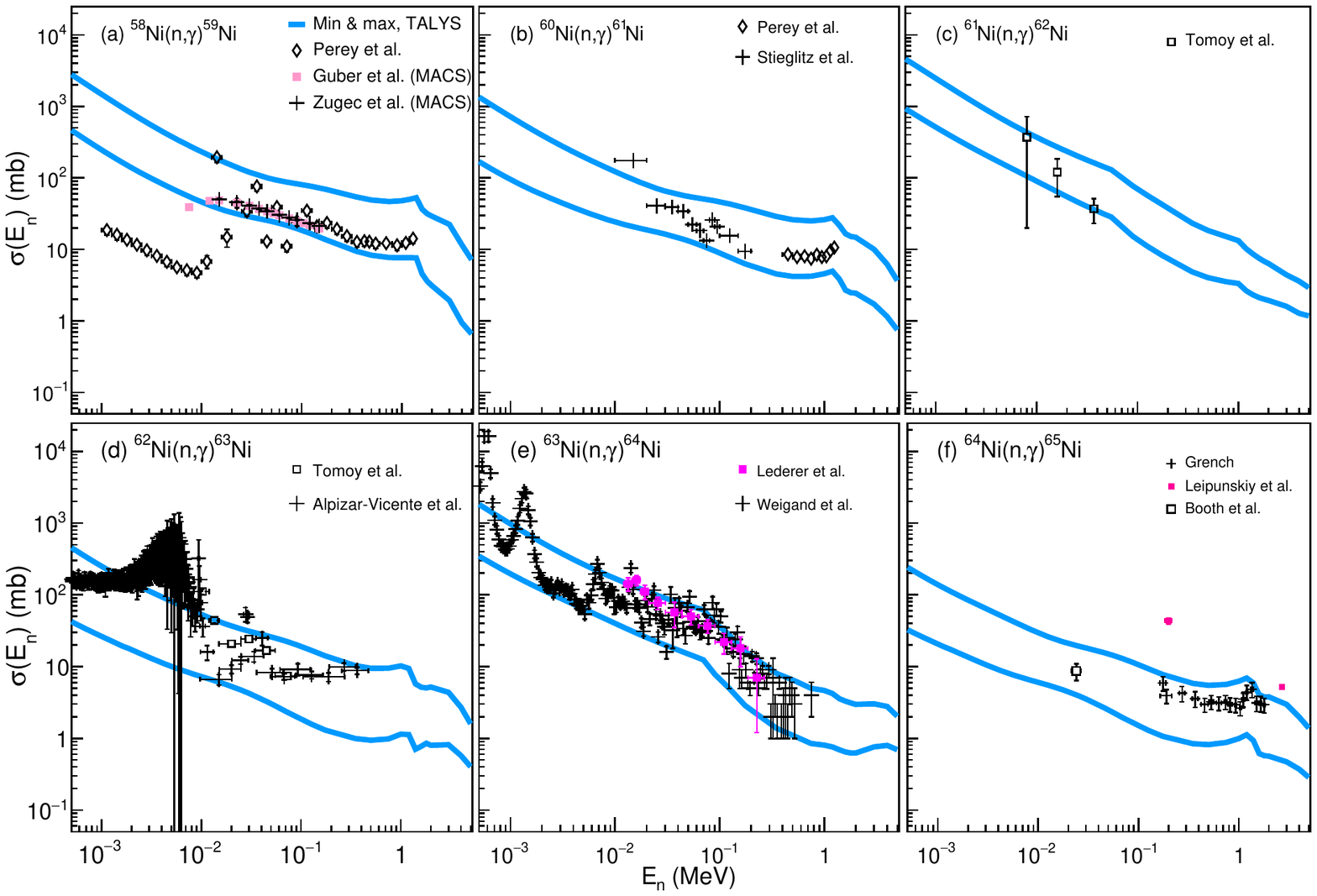}
\caption{(Color online) Same as Fig.~\ref{fig:xsec_Fe_Co} for
(a) $^{58}$Ni(n,$\gamma$), data from Refs.~\cite{zugec2014,perey1993,guber2010}; 
(b) $^{60}$Ni(n,$\gamma$), data from Refs.~\cite{stieglitz1971,perey1982}; (c) $^{61}$Ni(n,$\gamma$), 
data from Ref.~\cite{tomyo2005}, (d) $^{62}$Ni(n,$\gamma$), data from Refs.~\cite{tomyo2005,alpizar2008}, 
(e) $^{63}$Ni(n,$\gamma$), data from Refs.~\cite{lederer2014,weigand2015}, and (f) $^{64}$Ni(n,$\gamma$), 
data from Refs.~\cite{grench1965,leipunskiy1958,booth1958}.}
\label{fig:xsec_Ni}
\end{center}
\end{figure*}
\begin{figure*}[!hbt]
\begin{center}
\includegraphics[clip,width=1.2\columnwidth]{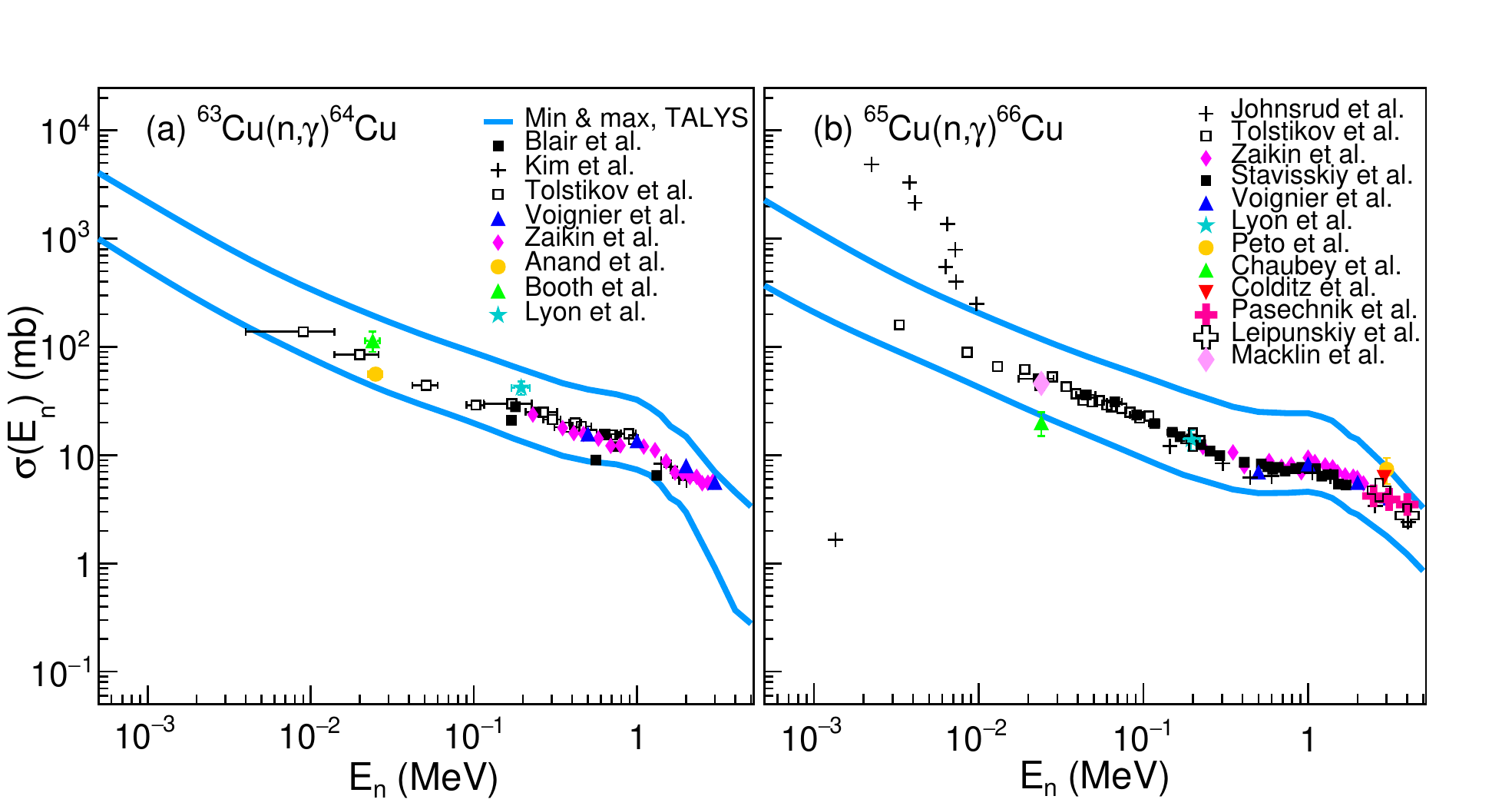}
\caption{(Color online) Same as Fig.~\ref{fig:xsec_Fe_Co} for (a) $^{63}$Cu(n,$\gamma$), data from 
Refs.~\cite{blair1944,voignier1992,tolstikov1966,zaikin1968,kim2007,anand1979,booth1958,lyon1959}; 
(b) $^{65}$Cu(n,$\gamma$), data from 
Refs.~\cite{johnsrud1959,voignier1992,zaikin1968,leipunskiy1958,pasechnik1958,stavisskiy1961,tolstikov1964,lyon1959,macklin1957,peto1967,chaubey1966,colditz1968}.}
\label{fig:xsec_Cu}
\end{center}
\end{figure*}
\begin{figure*}[!htb]
\begin{center}
\includegraphics[clip,width=1.7\columnwidth]{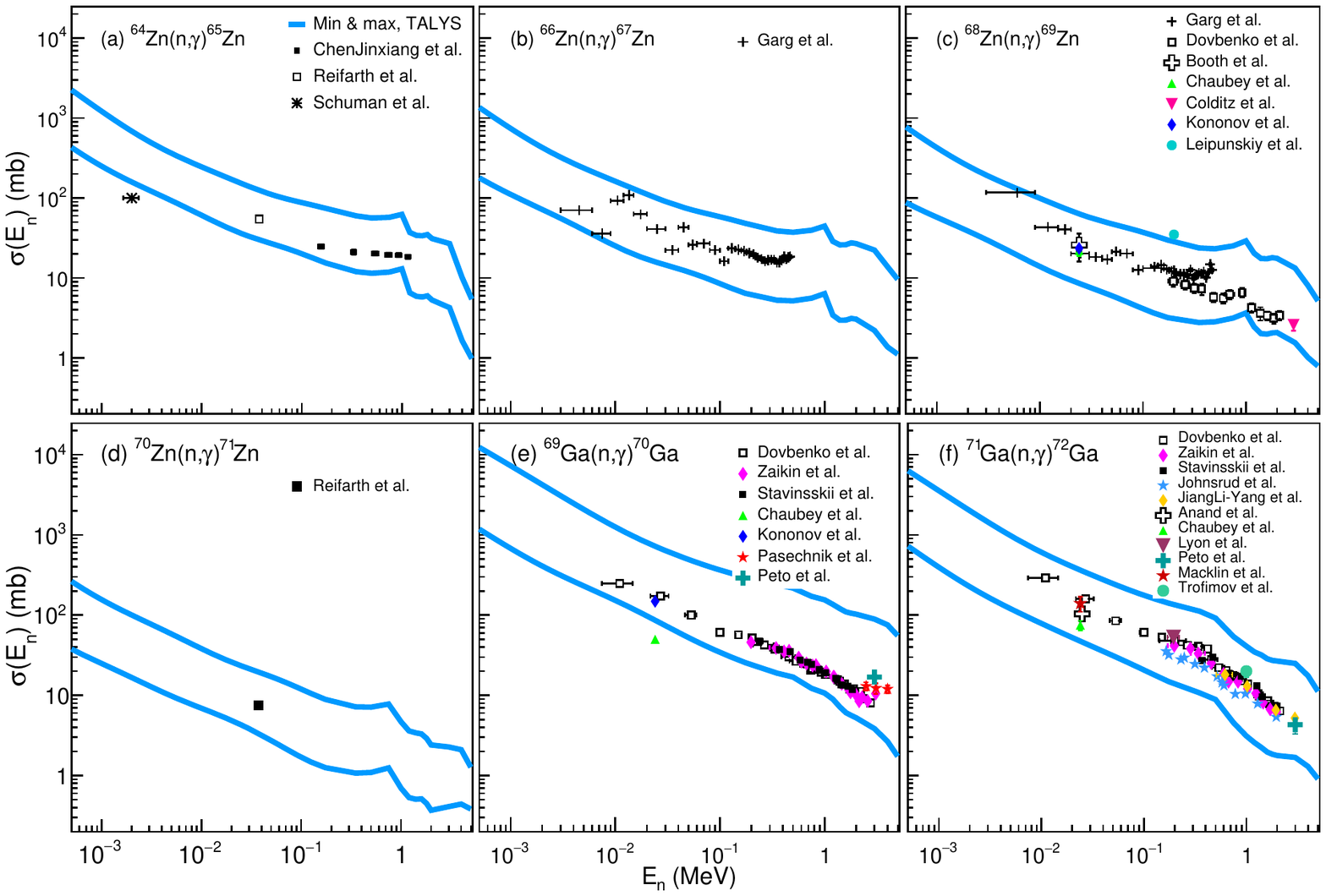}
\caption{(Color online) Same as Fig.~\ref{fig:xsec_Fe_Co} for (a) $^{64}$Zn(n,$\gamma$), data from 
Refs.~\cite{jinxiang1995,reifarth2012,schuman1970}; 
(b) $^{66}$Zn(n,$\gamma$), data from 
Ref.~\cite{garg1981}; (c) $^{68}$Zn(n,$\gamma$), data from 
Refs.~\cite{leipunskiy1958,booth1958,chaubey1966,colditz1968,garg1982,dovbenko1973,kononov1958};
(d) $^{70}$Zn(n,$\gamma$), data from Ref.~\cite{reifarth2012};
(e) $^{69}$Ga(n,$\gamma$), data from Refs.~\cite{dovbenko1969,zaikin1971,stavisskii1959,chaubey1966,kononov1958,pasechnik1958,peto1967};
(f) $^{71}$Ga(n,$\gamma$), data from Refs.~\cite{johnsrud1959,dovbenko1969,zaikin1971,stavisskii1959,liyang2012,anand1979,chaubey1966,lyon1959,macklin1957,peto1967,trofimov1987}.}
\label{fig:xsec_Zn_Ga}
\end{center}
\end{figure*}

\begin{figure*}[!hbt]
\begin{center}
\includegraphics[clip,width=1.6\columnwidth]{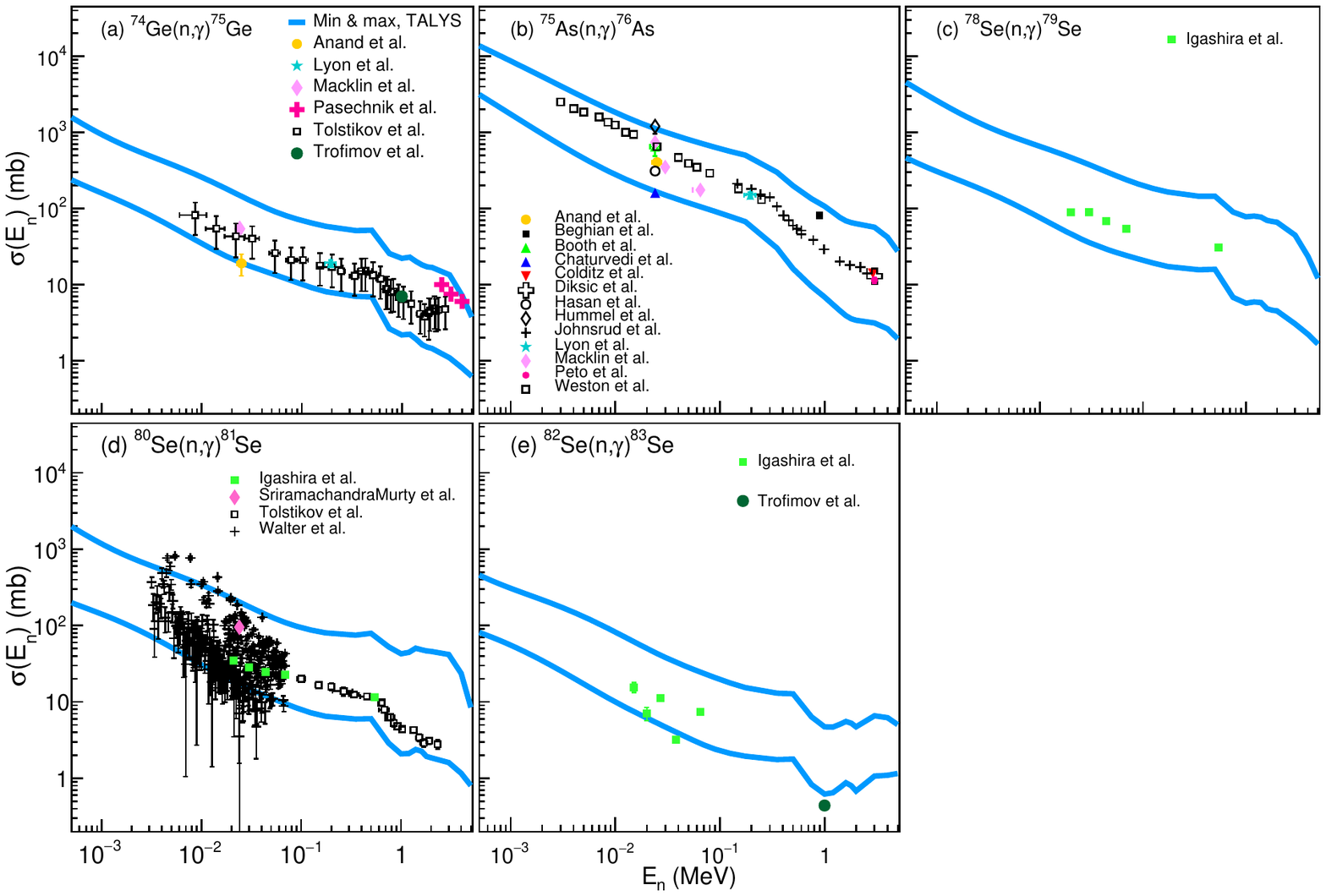}
\caption{(Color online) Same as Fig.~\ref{fig:xsec_Fe_Co} for (a) $^{74}$Ge(n,$\gamma$), data from 
Refs.~\cite{tolstikov1967,pasechnik1958,anand1979,lyon1959,macklin1957,trofimov1987}; 
(b) $^{75}$As(n,$\gamma$), data from Refs.~\cite{johnsrud1959,anand1979,booth1958,lyon1959,macklin1957,peto1967,colditz1968,beghian1949,chaturvedi1970,diksic1970,hasan1968,hummel1951,macklin1963,weston1960}; 
(c) $^{78}$Se(n,$\gamma$), data from Ref.~\cite{igashira2011};
(d) $^{80}$Se(n,$\gamma$), data from Refs.~\cite{tolstikov1967,igashira2011,sriramachandramurty1973,walter1984};
(e) $^{82}$Se(n,$\gamma$), data from Refs.~\cite{trofimov1987,igashira2011}.}
\label{fig:xsec_Ge_As_Se}
\end{center}
\end{figure*}

We have employed the nuclear-reaction code TALYS-1.8~\cite{TALYS_16,koning12} to calculate radiative neutron-capture
cross sections of Fe, Co, Ni, Cu, Zn, Ga, Ge, As, and Se isotopes where direct measurements are available 
from the EXFOR database~\cite{EXFOR}. A list of considered nuclei is given in Tab.~\ref{tab:list_stable}, with 
references to direct measurements for each case. 

The TALYS code treats both direct, pre-equilibrium and compound 
reactions, where the compound part is often the most important one for radiative neutron capture~\cite{Xu2014}, 
provided that the level density at the neutron binding energy of the compound nucleus is fairly high. 
Roughly, the compound-nucleus part of the reaction cross section for radiative neutron capture is given by
\begin{equation}
\sigma_{n\gamma} \propto \mathcal{T}_n(E_n) \rho(E_n+S_n) \mathcal{T}(E_\gamma),
\label{eq:cross}
\end{equation}
where $\mathcal{T}_n$ is the neutron transmission coefficient for the target-plus-neutron
system determined by the n-OMP, 
$\rho(E_n+S_n)$ is the NLD in the compound
system at the
neutron separation energy $S_n$ plus the incoming neutron energy $E_n$, and $\mathcal{T}(E_\gamma)$
is the $\gamma$-ray transmission coefficient directly proportional to the
$\gamma$SF, $f(E_\gamma)$, by 
\begin{equation}
\mathcal{T}(E_\gamma) = 2\pi E_\gamma^{2L+1} f(E_\gamma).
\end{equation}
Here, $L$ gives the multipolarity of the $\gamma$ transition, which is dominantly dipole
($L=1$) at excitation energies close to $S_n$. 

For the NLD, TALYS-1.8 has six models implemented: 
\begin{itemize}
\item[1.] {The combined constant-temperature~\cite{ericson1959} plus Fermi gas model~\cite{bethe1936} as first presented in Ref.~\cite{gilbert1965}, with
parameters as given in the TALYS manual (input keyword \textit{ldmodel 1});}
\item[2.] {The back-shifted Fermi gas (BSFG) model~\cite{gilbert1965} with
parameters as given in the TALYS manual ( keyword \textit{ldmodel 2});}
\item[3.] {The generalized superfluid (GSF) model~\cite{ignatyuk1979,ignatyuk1993} (keyword \textit{ldmodel 3});}
\item[4.] {Calculations based on the Hartree-Fock-Bardeen-Cooper-Schrieffer (HF-BCS) 
approach~\cite{demetriou2001} (keyword \textit{ldmodel 4});}
\item[5.] {Calculations within the Hartree-Fock-Bogoliubov (HFB) plus combinatorial framework~\cite{goriely2008}
(keyword \textit{ldmodel 5});}\\
\item[6.] {A temperature-dependent HFB plus combinatorial
level density~\cite{hilaire2012} (keyword \textit{ldmodel 6}).}
\end{itemize}
Note that some of these models have default parameters that are tuned to known data such as
the discrete levels at low excitation energy and neutron-resonance spacings at the 
neutron binding energy. For example, the microscopic models 4--6 are adjusted to 
these data, if available, by a fit function including an excitation-energy shift and a slope 
correction. In all cases, we have used the default parameters as implemented in TALYS. 
Obviously, for very unstable nuclei such data are not available and so the original
calculation is used. For more details, see the corresponding references and/or the TALYS-1.8 manual.

For the electric dipole ($E1$) component of the $\gamma$SF, eight models are available: 
\begin{itemize}
\item[1.]{The generalized Lorentzian model~\cite{kopecky_uhl_1990} (keyword \textit{strength 1});}
\item[2.]{The standard (Brink-Axel) Lorentzian model~\cite{brink1955,axel1962} (keyword \textit{strength 2});}
\item[3.]{Microscopic calculations within the quasi-particle random-phase approximation (QRPA) for excitations 
on top of HF-BCS calculations~\cite{goriely2002} (keyword \textit{strength 3});}
\item[4.]{QRPA calculations on top of HFB calculations~\cite{goriely2004} (keyword \textit{strength 4});}
\item[5.]{The hybrid model of Goriely~\cite{goriely1998} (keyword \textit{strength 5});}
\item[6.]{QRPA calculations as in Ref.~\cite{goriely2004}, but on top of temperature-dependent HFB calculations as in Ref.~\cite{hilaire2012} (keyword \textit{strength 6});}
\item[7.]{QRPA calculations combined with temperature-dependent relativistic mean-field calculations~\cite{daoutidis2010} (keyword \textit{strength 7});}
\item[8.]{QRPA calculations on top of deformed-basis HFB calculations using the Gogny force~\cite{martini2016} (keyword \textit{strength 8}).}
\end{itemize}
For the magnetic dipole ($M1$) strength, we use the default TALYS approach, which corresponds to an $M1$ spin-flip
resonance with standard parameterization as of Ref.~\cite{RIPL3}. Note that we specifically required
that no normalization of the $\gamma$SF to known average, radiative widths $\left< \Gamma_\gamma \right>$
was done (keyword \textit{gnorm 1.}). In this way we could assess the predictive power of the models and avoiding
a phenomenological re-scaling to match data.  

Finally, for the n-OMP, we have used the following two approaches:
\begin{itemize}
\item[1.]{The phenomenological, global parameterization of 
Koning and Delarouche~\cite{koning03} (keyword \textit{localomp n});}
\item[2.]{The semi-microscopic optical potential of the Jeukenne-Lejeune-Mahaux 
(JLM) type~\cite{bauge2001} (keyword \textit{jlmomp y}).}
\end{itemize}
In most cases, the JLM potential is within 20-30\% of the Koning-Delarouche potential. We also note that 
the JLM potential is typically lower than the Koning-Delarouche one for $E_n \approx 1 - 100$ keV. However, for
higher neutron energies, typically above 1 MeV, there is in some cases a cross-over and the JLM potential gives a 
higher ($n,\gamma$) cross section. 

The resulting lower and upper TALYS-1.8 predictions, together with available $(n,\gamma)$ 
cross-section data, are 
shown in Figs.~\ref{fig:xsec_Fe_Co}--\ref{fig:xsec_Ge_As_Se} for the Fe, Co, Ni, Cu, Zn, Ga, Ge, As, and Se isotopes. Overall, the TALYS lower and upper cross sections are of the order of a 
factor $\sim 5-10$. It is interesting to note that the spread in the \textit{measured} cross sections are
in some cases even larger than a factor of 10; more specifically, this is seen in
the $^{75}$As($n,\gamma$) cross section at $E_n \simeq 24$ keV in Fig.~\ref{fig:xsec_Ge_As_Se}b. 
It is also obvious from Table~\ref{tab:list_stable} that in general, the standard Lorentzian model is always
involved in the maximum TALYS prediction, while the generalized Lorentzian is often a component in the minimum one.
On the level-density side the picture is not as clear, although the constant-temperature plus Fermi-gas model is most frequently giving the
lowest cross section, while the BSFG and GSF models are often predicting the highest cross section.
Regarding the n-OMP, the JLM potential gives typically a lower cross section than the 
Koning-Delarouche potential as mentioned above; however, there are cases where the opposite is true, 
in particular $^{74}$Ge($n,\gamma$)$^{75}$Ge and $^{80}$Se($n,\gamma$)$^{81}$Se. 

Further, we see that although the upper and lower limits typically wrap around the data,
there are some notable exceptions, such as the $^{58}$Ni($n,\gamma$) cross section for 
neutron energies below $\approx 10$ keV, and the resonance region in the 
$^{62,63}$Ni($n,\gamma$) and $^{80}$Se($n,\gamma$) cross sections below $\approx 10$ keV. 
It is maybe not so surprising that the resonance region is not well described,
and the "dip" in the  $^{58}$Ni($n,\gamma$) might indicate that the compound-nucleus
part of the cross section is much smaller (about a factor of 10) than what TALYS
predicts. This might hint to the necessity of an even more sophisticated treatment of the direct and 
pre-equilibrium components. 

Moreover, we note that the minimum TALYS prediction is consistently lower than the majority of the
data, while the maximum cross section is surprisingly close to the experimental data for 
the case $^{63}$Ni($n,\gamma$). This underlines the fact that the 
phenomenological models are not able to properly catch the underlying nuclear structure
explaining the measured cross section. Also, since the calculation basically involves a folding of the
NLD, $\gamma$SF, and n-OMP (see Eq.~(\ref{eq:cross}), it is not straightforward to test
these models based on a cross-section measurement alone; one would need data to probe the
NLD, $\gamma$SF, and n-OMP separately to be able to identify the correct input for 
calculating the cross section.

From the comparison of the minimum and maximum TALYS calculations with data for nuclei
in the valley of stability, it is quite striking that the predictive power of these 
frequently used models is not better than typically a factor $\sim 5-10$. It is in fact rather
depressing, as the desired uncertainty for ($n,\gamma$) reaction rates involved in the $r$-process
is less than a factor of 10~\cite{liddick2016}. 

Now that the $r$-process is shown to take place in neutron star mergers,  prompt ejecta 
from compact mergers (\textit{e.g.},~\cite{just15,foucart2014,Kasen2017}) are believed to be rather cold and
an ($n,\gamma$)--($\gamma,n$) 
equilibrium will never be reached~\cite{arnould2007}. 
Hence, the ($n,\gamma$) rates play a crucial role. 
In addition, post-merger neutrino and viscously driven outflows~\cite{just15} are believed to provide a hot, high-entropy environment, where ($n,\gamma$) reaction rates would mainly be important
in the freeze-out phase of the ($n,\gamma$)--($\gamma,n$) equilibrium (\textit{e.g.}, Ref.~\cite{mumpower2016}).
In either case, it is clear that there is an urgent need to 
pin down the underlying quantities and improve the model predictions. 
Considering also that the models should be able to handle exotic nuclei away from stability,
the only viable solution seems to be microscopic approaches~\cite{arnould2007} -- both  for
the NLD, $\gamma$SF and n-OMP as well as for the description of the reaction mechanisms.

In the following, we examine the uncertainties 
in predicted ($n,\gamma$) reaction rates for neutron-rich nuclei of the elements
Fe, Co, Ni, Cu, Zn, Ga, Ge, As, and Se. In particular, we focus on the cases with the highest
sensitivity measures according to Ref.~\cite{surman2014}.

\section{Reaction rate calculations for neutron-rich nuclei}
\label{sec:exotic}
\begin{figure*}[!hbt]
\begin{center}
\includegraphics[clip,width=1.6\columnwidth]{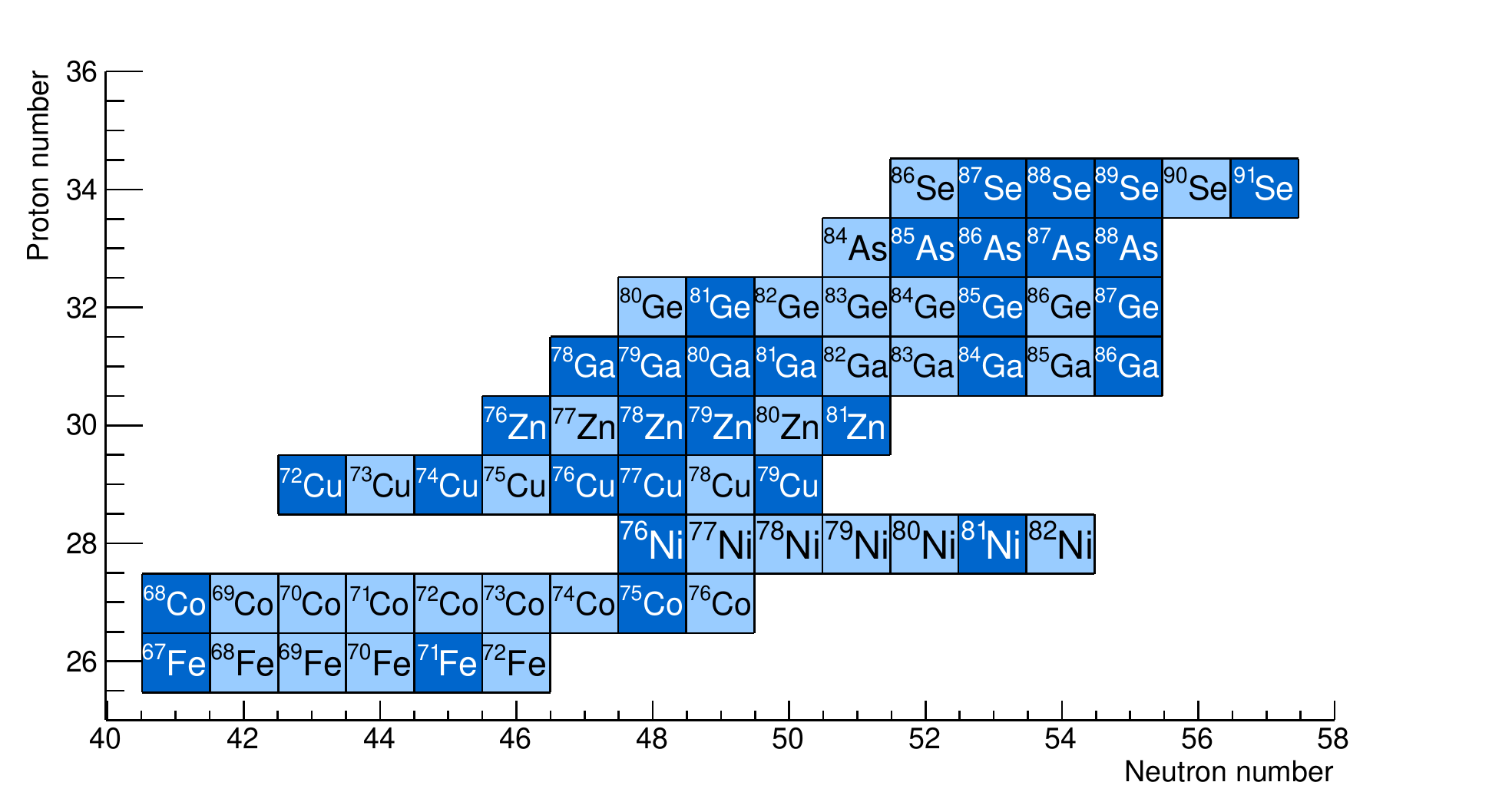}
\caption{(Color online) Considered cases for investigating the variation in ($n,\gamma$) reaction rates.
The cases with sensitivity measure higher than 10 according to Ref.~\cite{surman2014} are marked in dark blue.}
\label{fig:interest_cases}
\end{center}
\end{figure*}
The authors of Ref.~\cite{surman2014}, as shown in their Fig.~7 and Table~II, have identified certain ($n,\gamma$) reaction rates 
to be of particular importance, even for a large variety of astrophysical conditions. 
In their study, the ($n,\gamma$) reaction rates were taken from the JINA REACLIB 
library~\cite{JINA-REACLIB}
and scaled up and down with a factor of 10, spanning an uncertainty range of a factor of 100
relative to the JINA REACLIB recommended rates.  
Surman \textit{et al.}~\cite{surman2014} also define a sensitivity measure given by 
\begin{equation}
F = 100 \times \sum_{A} \left|  X(A) - X_{\mathrm{baseline}}(A) \right|,
\end{equation}
where $X_{\mathrm{baseline}}(A)$ are the final mass fractions for the baseline nuclear-network 
calculations for a given set of astrophysical conditions (entropy per baryon, dynamic timescale, 
and the electron fraction). Correspondingly, $X(A)$ represent the final mass fractions for 
the simulation where the ($n,\gamma$) rates are changed up or down by a factor of 10. In this
way, combining the results from 55 different astrophysical trajectories that showed significant
sensitivity to neutron-capture rates, cases with particular importance for this 
range of trajectories (i.e. astrophysical conditions) were identified. 

In this section, we investigate what uncertainties TALYS predicts using all possible combinations
of level-density and $\gamma$SF models for the rates of the considered moderately neutron-rich nuclei shown in Fig~\ref{fig:interest_cases}. 
The "target" nuclei marked in dark blue are the ones with sensitivity measure higher than 10 according to Ref.~\cite{surman2014}. 
We also vary the n-OMP as well as the mass models available in TALYS, which are: 
\begin{itemize}
\item[1.]{The Duflo-Zuker mass formula~\cite{duflo1995} (keyword \textit{massmodel 0});}
\item[2.]{The M\"{o}ller table~\cite{moller1995} (keyword \textit{massmodel 1});}
\item[3.]{The Skyrme-HFB calculations described in Ref.~\cite{goriely2009},
where an unprecedented rms deviation with experimental masses for a mean-field approach of $< 0.6$ MeV is achieved (keyword \textit{massmodel 2});}
\item[4.]{The Gogny-D1M HFB calculations of Ref.[REF] (keyword \textit{massmodel 3}).}
\end{itemize}

\begin{figure}[!hbt]
\begin{center}
\includegraphics[clip,width=\columnwidth]{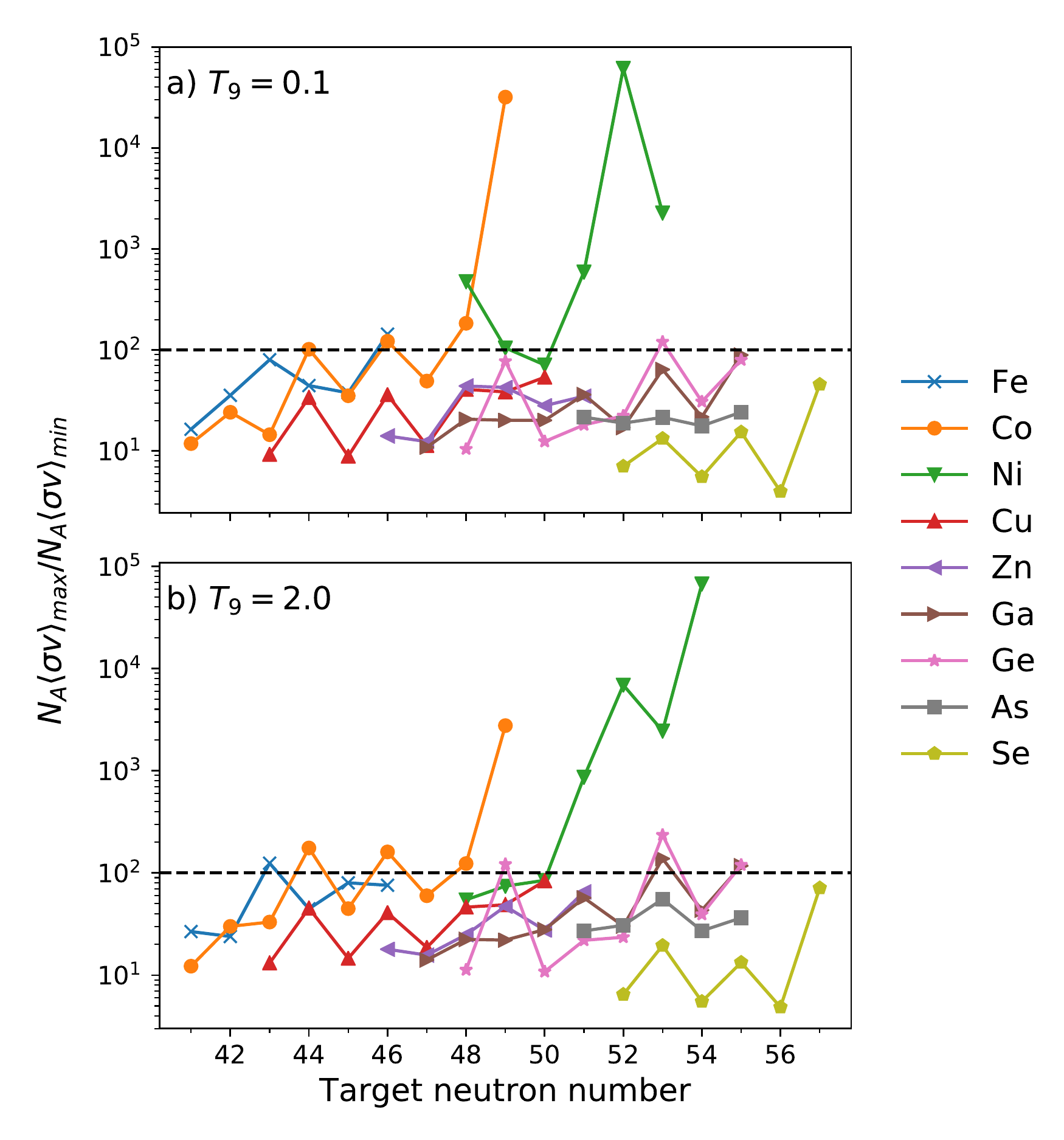}
\caption{(Color online) 
Ratios of the maximum and minimum ($n,\gamma$) reaction rates
calculated with TALYS.
The ratios are taken at temperatures (a) $T = 0.1$ GK and (b) $T = 2$ GK in the astrophysical environment.
The dashed line indicates the considered variation in Ref.~\cite{surman2014} of a factor 100. In a) the data point of the $^{82}$Ni($n,\gamma$) ratio is not included due to the TALYS minimum rate being zero. }
\label{fig:min_max_TALYS}
\end{center}
\end{figure}

In Fig.~\ref{fig:min_max_TALYS}, the ratio of the maximum and minimum ($n,\gamma$) reaction rates 
calculated with TALYS is shown for the cases in Fig.~\ref{fig:interest_cases}. We see that 
the ratio is typically a factor $\sim 10-100$, but there are cases that are even more extreme with a factor of $\sim 10^3 -10^5$ such as: a) $^{76}$Co($n,\gamma$) and $^{76,79,80,81}$Ni($n,\gamma$) for $T = 0.1$ GK and b) $^{76}$Co($n,\gamma$) and $^{79,80,81,82}$Ni($n,\gamma$) for $T=2$ GK. 
This means that the actual variation in reaction rates could be even larger than 
the factor of 100 considered in Ref.~\cite{surman2014}. For the particularly important 
cases with high sensitivity measure, we see that for $T_9 = 0.1$, 
$^{81}$Ni($n,\gamma$) varies with a factor of $\sim 2000$, $^{76}$Ni($n,\gamma$) varies with a factor of $\sim 500$ and $^{75}$Co($n,\gamma$) vary by approximately by a factor 200.
For higher temperatures of
$T_9 = 2.0$, $^{81}$Ni($n,\gamma$) varies by a factor $\sim 2000$ and $^{85}$Ge($n,\gamma$) vary by $\sim 200$. 
The data point of the $^{82}$Ni($n,\gamma$) rate at $T=0.1$ is not included in Fig.~\ref{fig:min_max_TALYS} due to the TALYS minimum rate being zero. 

\begin{figure}[!htb]
\begin{center}
\includegraphics[clip,width=1.\columnwidth]{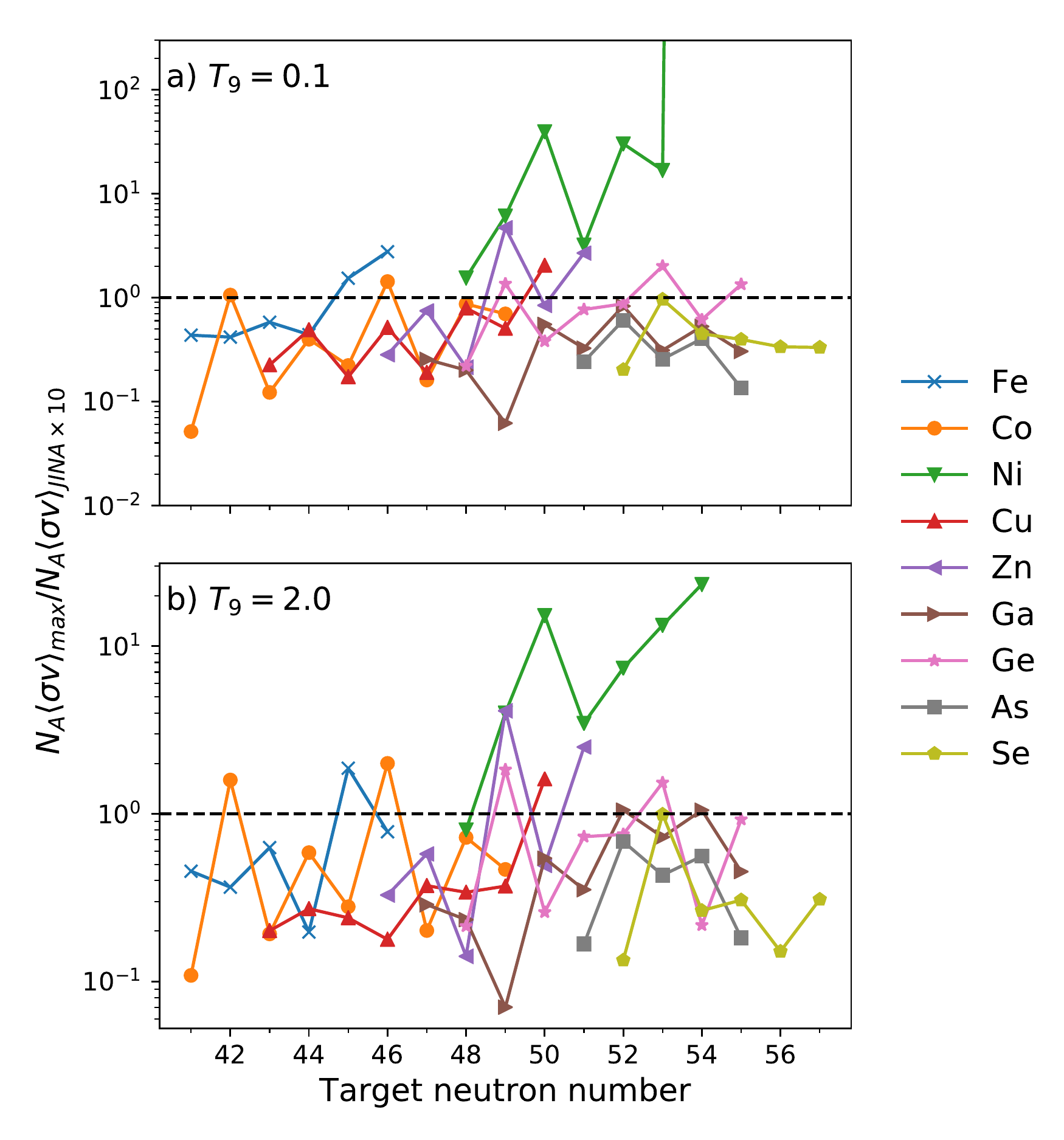}
\caption{(Color online) Ratios of the maximum ($n,\gamma$) reaction rates
calculated with TALYS versus the JINA REACLIB rates multiplied by a factor of 10 for the cases 
shown in Fig~\ref{fig:interest_cases}.
The ratios are taken at temperatures (a) $T = 0.1$ GK and (b) $T = 2$ GK in the astrophysical environment.}
\label{fig:max_TALYS_JINA}
\end{center}
\end{figure}

Further, we compare the TALYS results with the JINA REACLIB rates for the cases of interest (see Fig. \ref{fig:interest_cases}). 
In Fig.~\ref{fig:max_TALYS_JINA}, the ratio of the \textit{maximum} TALYS rates over the 
JINA REACLIB rates multiplied with a factor 10 are shown, i.e. we are comparing with
the upper limit used in Ref.~\cite{surman2014}. 

In general, the maximum TALYS rates are smaller than or similar to the 
JINA REACLIB rates multiplied by 10. At $T_9=0.1$, for the cases $^{78,80,81}$Ni($n,\gamma$) the TALYS maximum rates are a factor $\sim 400$, $\sim300$ and $\sim160$ respectively, bigger than the JINA REACLIB rates. However, of these three cases only 
$^{81}$Ni($n,\gamma$) is considered important for the final abundances according to~\cite{surman2014}. 

Another extreme case is the $^{82}$Ni($n,\gamma$) reaction 
(off-scale in Fig.~\ref{fig:max_TALYS_JINA}a), which for the default 
JINA REACLIB rate is practically zero ($\sim 10^{-35}$ cm$^3$ $s^{-1}$ mol$^{-1}$). 
That said, the prediction of this rate is
extremely sensitive to which mass model is used; the $Q$-value varies from 
$-0.58$ MeV (FRDM mass input) to $1.09$ MeV (ETSFIQ mass input) -- 
for the latter the JINA REACLIB rate is $\approx 700$ cm$^3$ $s^{-1}$ mol$^{-1}$.

For the most important cases (marked dark blue in Fig.~\ref{fig:interest_cases}), 
we see that the $^{68}$Co($n,\gamma$) and $^{80}$Ga($n,\gamma$) maximum TALYS rate is much smaller than the 
JINA REACLIB rate by about a factor of $\sim15$ and $\sim 20$ respectively. On the other hand, the maximum TALYS rates for the sensitive cases of
$^{79}$Cu($n,\gamma$), $^{79,81}$Zn($n,\gamma$), $^{85}$Ge($n,\gamma$) are less than a factor $\sim 10$ larger than the corresponding JINA REACLIB rates, 
while the maximum TALYS rate of $^{81}$Ni($n,\gamma$) is a factor $\sim 20$ larger than the corresponding JINA REACLIB rates.  

For a higher temperature of the astrophysical environment with $T_9 = 2.0$, most of the rates seem to less than 10 times smaller or larger than the maximum JINA rates used in Ref.~\cite{surman2014}. The largest deviations of the sensitive rates are in the the maximum TALYS rates of $^{81}$Ni($n,\gamma$) that are around $\sim 15$ times larger than the JINA REACLIB rates and for $^{68}$Co($n,\gamma$) and $^{80}$Ga($n,\gamma$) that are 10 times smaller than the maximum JINA rate considered in Ref.~\cite{surman2014}.
\begin{figure}[!htb]
\includegraphics[clip,width=\columnwidth]{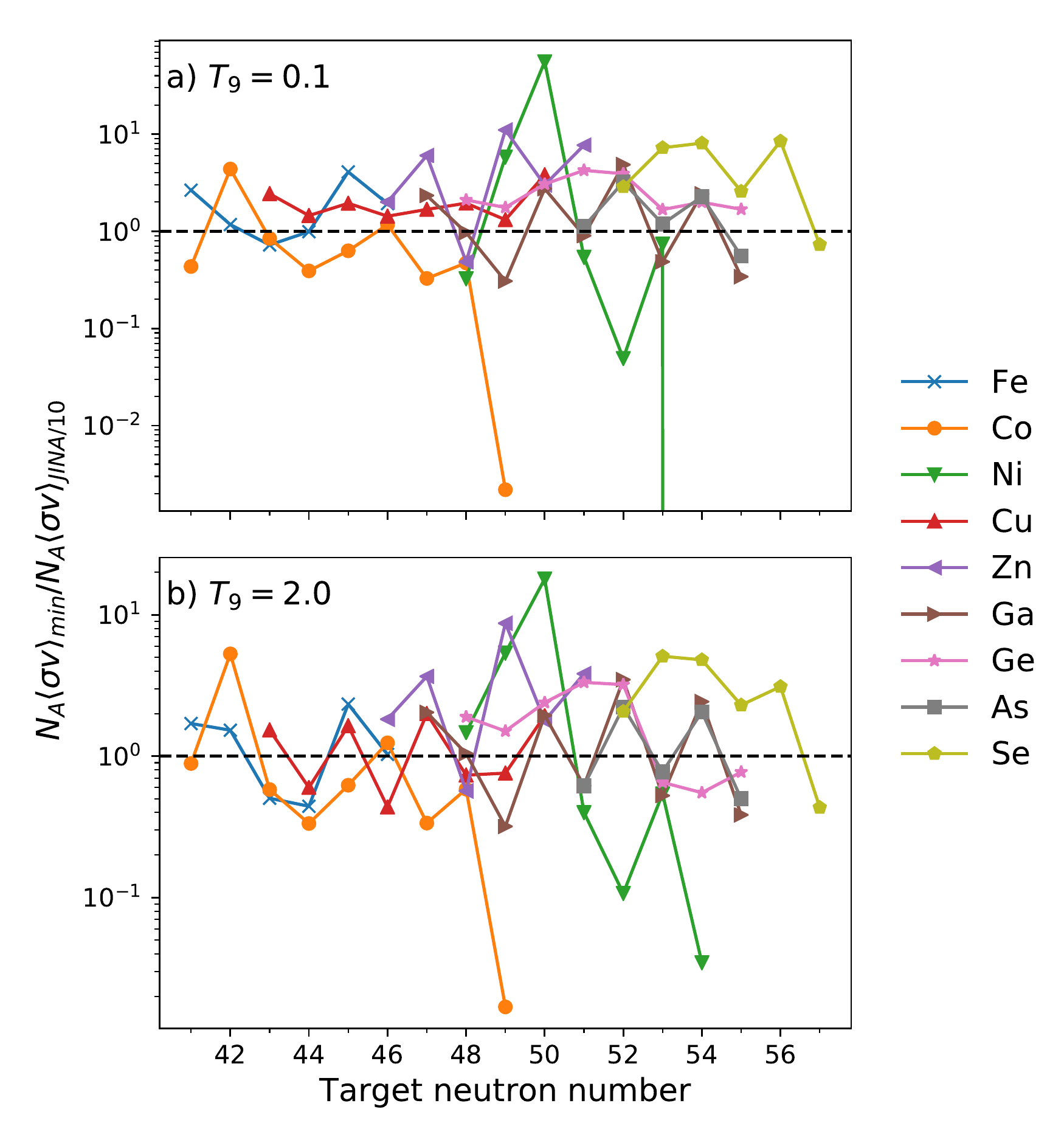}
\caption{(Color online) Ratios of the minimum ($n,\gamma$) reaction rates
calculated with TALYS versus the JINA REACLIB rates divided by a factor of 10 for the cases 
shown in Fig~\ref{fig:interest_cases}.
The ratios are taken at temperatures (a) $T = 0.1$ GK and (b) $T = 2$ GK in the astrophysical environment.}
\label{fig:min_TALYS_JINA}
\end{figure}

Next, we compare the minimum TALYS rates with the JINA REACLIB rates divided
by a factor of 10, i.e. the lower limit considered in~\cite{surman2014}. 
This ratio is shown in Fig.~\ref{fig:min_TALYS_JINA} where we see a large variation in the estimated minimum rates.
For $T_9 = 0.1$, the most extreme cases are $^{76}$Co($n,\gamma$) and $^{80}$Ni($n,\gamma$) where the lower rates from TALYS are a factor of $\sim 10$ and $\sim 1000$ smaller than the lower JINA REACLIB rates. 
In addition, the case $^{82}$Ni($n,\gamma$) is completely off-scale in Fig. \ref{fig:min_TALYS_JINA}) because of 
the nearly zero minimum TALYS rate.
On the other hand, the TALYS minimum rate for $^{78}$Ni($n,\gamma$) is more than a factor of 10 larger than the JINA REACLIB lower rate. 

At $T_9 = 2.0$, the overall trend is a spread around the ratio of 1, the TALYS minimum rates are about equal to the JINA REACLIB rates within a factor of $\approx 10$. In particular $^{76}$Co($n,\gamma$) and $^{80,82}$Ni($n,\gamma$) have a factor of $\sim 10-100$ lower rates while $^{78}$Ni($n,\gamma$) have a factor of $\sim 200$ higher rate than the JINA REACLIB reaction rate. None of the rates with high sensitivity measure considered in Ref.~\cite{surman2014} have a deviation of more than a factor of 10 up or down at this astrophysical temperature. 

\begin{figure}[!htb]
\begin{center}
\includegraphics[clip,width=0.95\columnwidth]{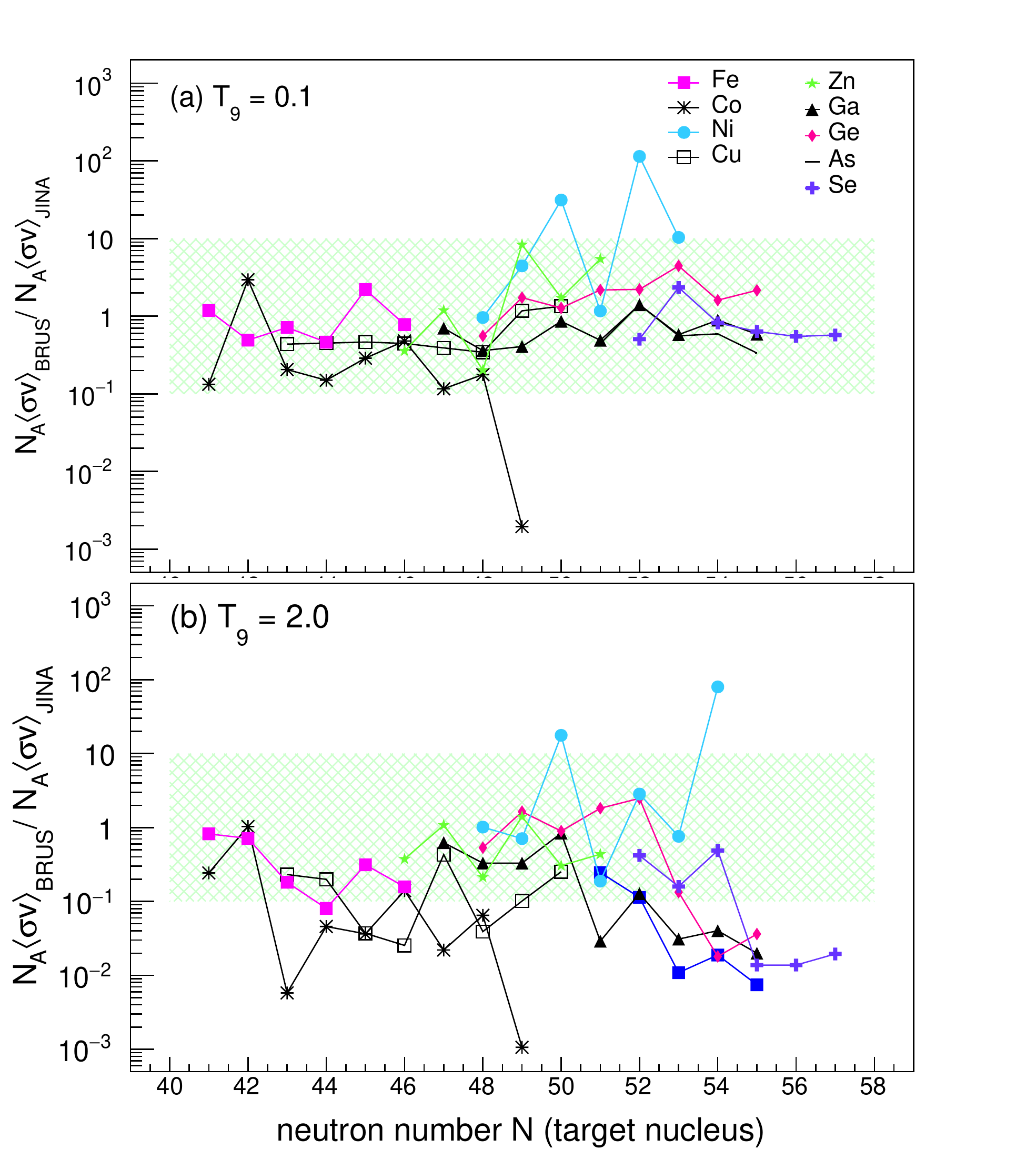}
\caption{(Color online) Ratios of BRUSLIB ($n,\gamma$) reaction rates
versus JINA REACLIB for the cases 
shown in Fig~\ref{fig:interest_cases}.
The ratios are taken at temperatures (a) $T = 0.1$ GK and (b) $T = 2$ GK in the astrophysical environment.
The shadowed band indicates a factor of 100 deviation (factor 10 up or down) as used in the sensitivity study by Ref.~\cite{surman2014}. In a) the data point of the $^{82}$Ni($n,\gamma$) ratio is not included due to the JINA REACLIB rate being extremely low.  
}
\label{fig:BRUSLIB_JINA}
\end{center}
\end{figure}
Now, we would like to investigate the variation between the JINA REACLIB and BRUSLIB
rates, since both these libraries are frequently used in $r$-process reaction network
calculations. The result is shown in Fig.~\ref{fig:BRUSLIB_JINA}. We see that for 
$T_9=0.1$, most of the reaction rates are within a factor of 100, except for the cases 
$^{76}$Co($n,\gamma$) and $^{78,80}$Ni($n,\gamma$). In addition, the point of the $^{82}$Ni($n,\gamma$) ratio is not included in Fig.~\ref{fig:BRUSLIB_JINA} due to the JINA REACLIB rate being close to zero.
For $T_9 = 2.0$ the situation is different, with many BRUSLIB rates being more than a factor 10 smaller than the JINA 
REACLIB ones. Again focusing on the cases with the highest sensitivity measure, 
the $^{75}$Co, $^{74,77}$Cu, $^{84,86}$Ga($n,\gamma$), $^{87}$Ge($n,\gamma$), $^{86,87,88}$As($n,\gamma$) 
and $^{89,91}$Se($n,\gamma$) BRUSLIB rates are up to a factor $\approx10^{-2}$ smaller than 
the JINA REACLIB predictions.  

Further, we consider the possible impact of an enhanced probability for emitting
low-energy $\gamma$ rays. At present, there is not so much experimental information 
on this feature, and we have chosen to implement it as a magnetic-dipole component
of the $\gamma$SF 
following several recent shell-model results~\cite{schwengner2013,brown2014,schwengner2017}.
Further, we make use of the phenomenological description of the $^{70}$Ni low-energy
$\gamma$SF~\cite{liddick2016,larsen2017} parameterized as
\begin{equation}
f_{\mathrm{up}}(E_\gamma) = C \exp\left[-\eta E_\gamma\right]
\label{eq:upbend}
\end{equation}
following e.g. Ref.~\cite{renstrom2016}. We apply $C=1.\cdot10^{-7}$ MeV$^{-3}$
and $\eta = 1.4$ MeV$^{-1}$ in accordance with the $^{70}$Ni data.
\begin{figure}[!htb]
\begin{center}
\includegraphics[clip,width=1.\columnwidth]{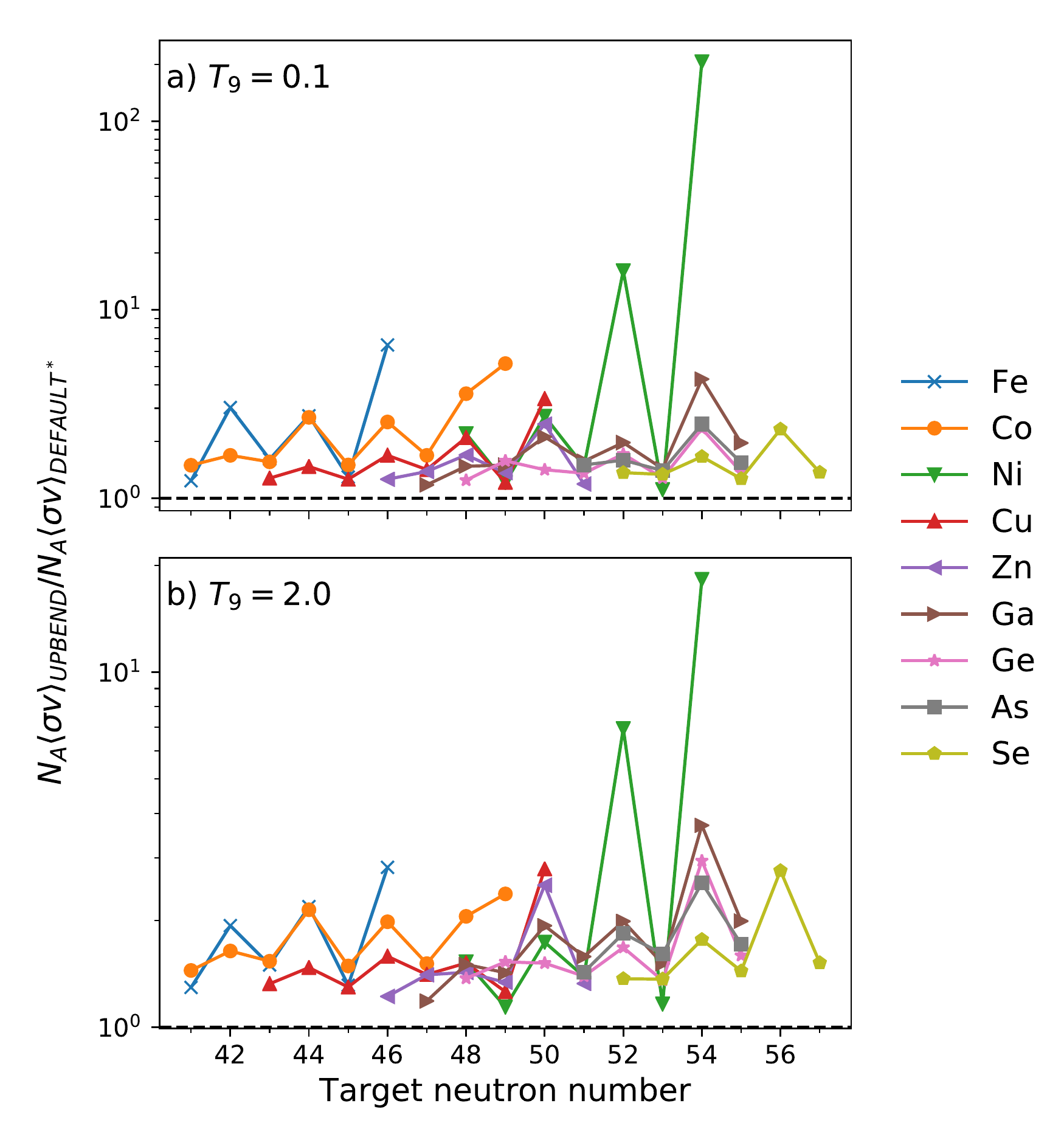}
\caption{(Color online) Ratios of ($n,\gamma$) reaction rates
with and without the low-energy enhancement (upbend) for the cases 
shown in Fig~\ref{fig:interest_cases} with the BRUSLIB-like TALYS input (bruslib*).
The ratios are taken at temperatures (a) $T = 0.1$ GK and (b) $T = 2$ GK in the astrophysical environment.}
\label{fig:upbend}
\end{center}
\end{figure}
\begin{figure}[!htb]
\begin{center}
\includegraphics[clip,width=1.\columnwidth]{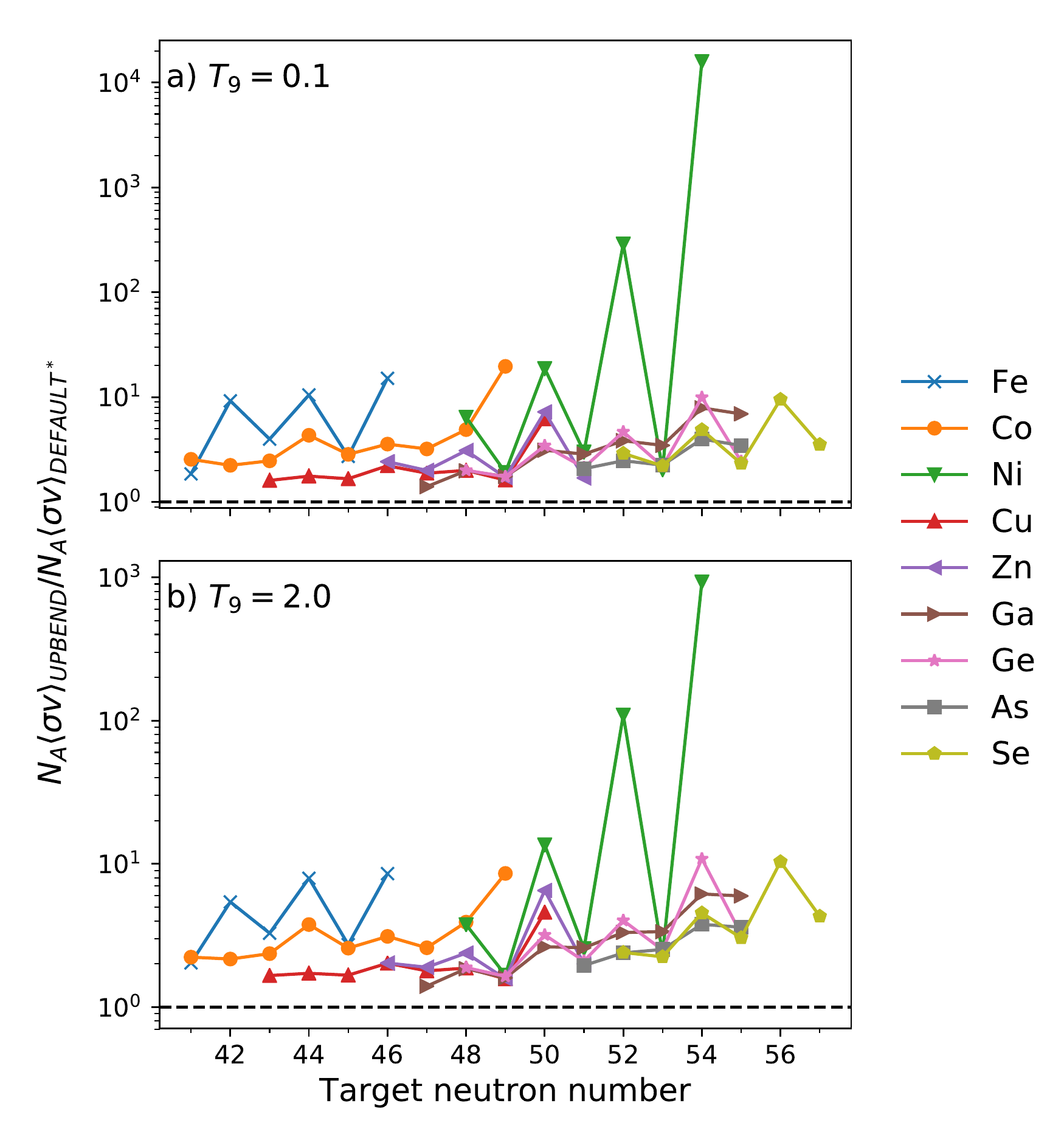}
\caption{(Color online) Ratios of ($n,\gamma$) reaction rates
with and without the low-energy enhancement for the cases 
shown in Fig~\ref{fig:interest_cases} with the TALYS default input.
The ratios are taken at temperatures (a) $T = 0.1$ GK and (b) $T = 2$ GK in the astrophysical environment.}
\label{fig:upbend_default}
\end{center}
\end{figure}
To see the influence of this low-energy enhancement on the reaction rates, we have 
added the $M1$ component in Eq.~(\ref{eq:upbend}) to the TALYS $\gamma$SF models. 
In particular, we have used input as close as possible to BRUSLIB~\cite{BRUSLIB}, i.e. 
the BSk17/HFB-17 mass model (\textit{massmodel 2}~\cite{goriely2009}) 
-- BRUSLIB uses a similar Skyrme force, BSk24, with slightly better rms deviation
for the masses of 0.55 MeV~\cite{gorielymass2013} -- the same NLD (\textit{ldmodel 5}~\cite{goriely2008}) and
$E1$ $\gamma$SF (\textit{strength 4}~\cite{goriely2004}). In Fig.~\ref{fig:upbend}, 
the ratio of the reaction rates with (upbend) and without (bruselib*) the low-energy enhancement is shown for the 
cases under study. We see that the low-energy enhancement does lead to an increase in the
reaction rates, but in general not more than a factor $\approx2-20$ for the nuclei studied here. The most extreme point is $^{82}$Ni($n,\gamma$) where the upbend rate is $\sim 200$ and $\sim 20$ times larger than the the rate without the enhancement for the astrophysical temperatures of $T=0.1$ and $T=2.0$ respectively. Hence, we see that the upbend always leads to an increase in the rates and represents another uncertainty combined with the uncertainties of varying the $\gamma$SF and NLD models.

We also compared the effect of including the low-energy enhancement in the
$\gamma$SF using the default input parameters in TALYS: \textit{ldmodel 1, strength 1, localomp n, massmodel 2}. 
The result is shown in Fig.~\ref{fig:upbend_default} and follow a similar trend to the BRUSLIB-like factors shown in Fig.~\ref{fig:upbend}. Again, the enhancement is in general leading to higher rates, but the TALYS default input result in a larger upbend over no upbend ratio. Here, the $^{80,82}$Ni the rates increase by a factor 
$\approx 300$ and $\approx 10^4$, respectively, for $T_9=0.1$, and $\approx 100$ and $\approx 1000$ for $T_9=2.0$.
The main reason for this is that the default $\gamma$SF model, 
the generalized Lorentzian model~\cite{kopecky_uhl_1990}
is dependent on the \textit{nuclear} temperature as $T_f \approx \sqrt{E_f/a}$, where $T_f$ is the temperature
at final excitation energy $E_f = E_i - E_\gamma$ and $a$ is the level-density parameter. When the neutron separation
energy goes down, so does $E_f$ and the temperature, and the $E1$ strength is reduced as well.
Hence, including a low-energy enhancement as parameterized in Eq.~(\ref{eq:upbend}) will strongly increase
the $\gamma$SF and consequently the ($n,\gamma$) reaction rate. Also, we note that the effect is most
significant for the Fe and Ni cases where the compound nucleus is odd, i.e. with a low neutron separation energy. 

Now, we would like to highlight some special cases where the reaction rates from the libraries and 
the TALYS predictions deviate in a particular way.  
\begin{figure}[!htb]
\begin{center}
\includegraphics[clip,width=\columnwidth]{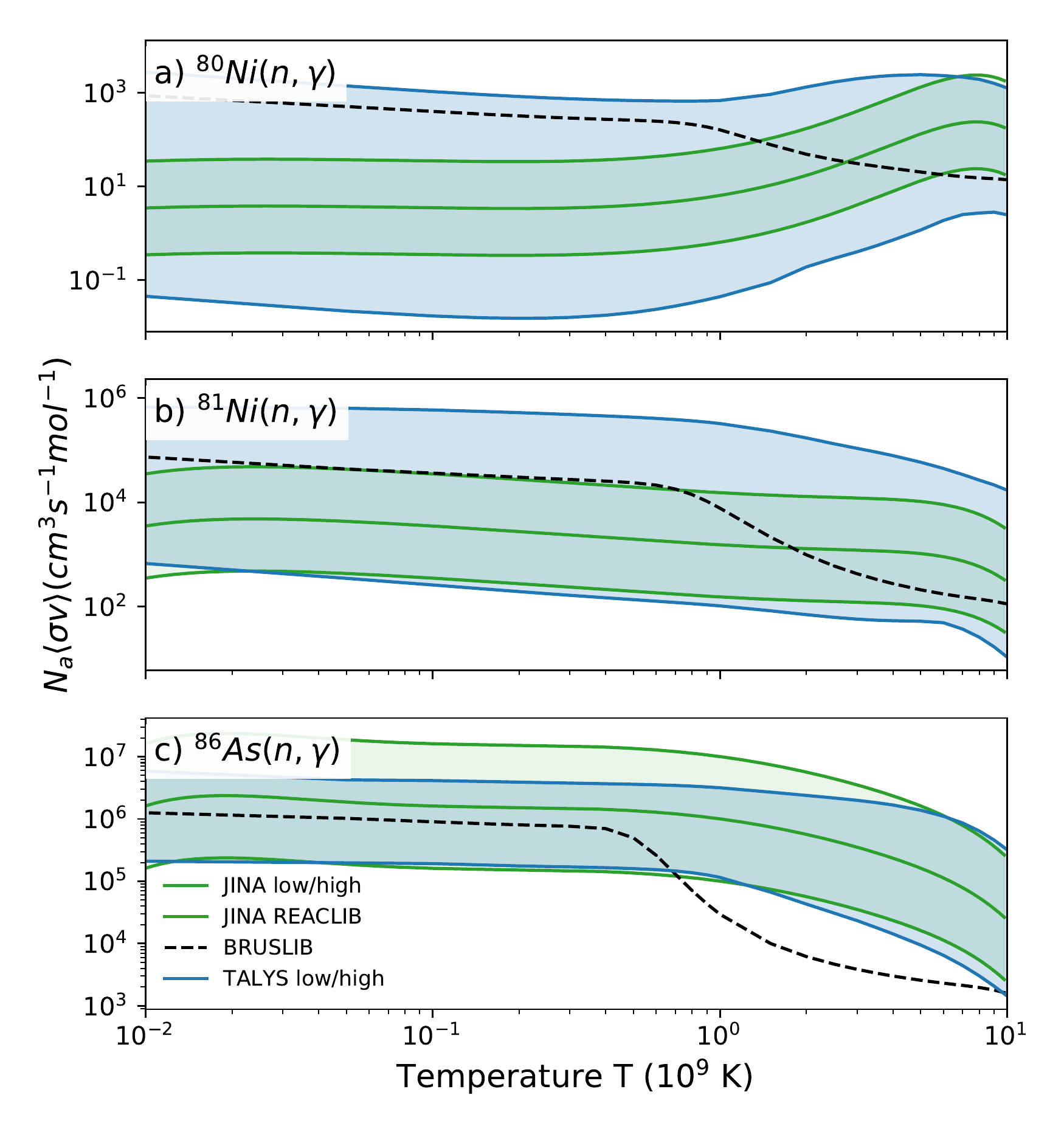}
\caption{(Color online) Reaction rates
for (a) $^{80}$Ni($n,\gamma$), (b) $^{81}$Ni($n,\gamma$), and  (c) $^{86}$As($n,\gamma$) (see text).
The shaded green band indicates the JINA REACLIB multiplied by a factor of 10 up and down and the shaded blue band indicates the TALYS minimum and maximum rates. }
\label{fig:specialcases}
\end{center}
\end{figure}
In Fig.~\ref{fig:specialcases}, the reaction rates of $^{80,81}$Ni($n,\gamma$) and
$^{86}$As($n,\gamma$) are shown. In all, we observe very different rates from
JINA REACLIB and TALYS as compared to BRUSLIB at high temperatures. Except for a very few crossings, the JINA REACLIB rates scaled with a factor 10 up or down are contained within the TALYS maximum and minimum rates. In general, the BRUSLIB rates have a different shape than the JINA REACLIB and TALYS rates. For instance we see that for $^{80}$Ni($n,\gamma$), the JINA REACLIB and TALYS rates increases significantly 
for $T_9>1$, with a peak around $T_9 \approx 7-8$, while in contrast, the BRUSLIB rate decreases
for $T_9>1$. 

One should keep in mind that there is a reported error in the logarithmic energy grid in one of the TALYS subroutines, 'partfunc.f', for TALYS versions 1.4--1.8 as described in the manual. 
This error is avoided in the current calculations by enforcing a linear energy grid for the astrophysical rate calculations (keyword \textit{equidistant y}). 
However, this error is very likely present in the BRUSLIB calculations, and could be the reason behind the different shapes of the rates at higher temperatures.




\section{Conclusions and outlook}
\label{sec:sum}
In this work we have investigated theoretical predictions of astrophysical 
($n,\gamma$) cross sections and reaction rates for Fe, Co, Ni, Cu, Zn, Ga, Ge, As, and Se isotopes, 
using the open-source reaction code TALYS. 
We have found that even for nuclei near the valley of stability, the spread in the predicted
cross sections is of the order of $5-10$. 

For neutron-rich nuclei, the variation between the 
lowest and highest ($n,\gamma$) reaction rates calculated with TALYS
is typically a factor $10-100$, reaching several orders of magnitude for some cases. 
We have followed Ref.~\cite{surman2014} and investigated some key reactions that may have a particularly large impact on the final abundances in the $A\sim 80$ region. 

Comparing the TALYS ($n,\gamma$) rates with the much-used JINA REACLIB and BRUSLIB
reaction rates, we have seen that there are several cases that display very large deviations, much more than the two orders of magnitude considered in Ref.~\cite{surman2014}. 
Further, the low-energy upbend in the $\gamma$ strength function has a non-negligible effect on the rates, although we note that the uncertainties due to other factors such as mass models and level densities may have much bigger effects on the rates. 

We conclude that there is a dire need for more data to constrain ($n,\gamma$) reaction rates in this mass region. New experimental techniques such as the surrogate method for neutron rich nuclei~\cite{kozub2012} and the beta-Oslo method~\cite{spyrou2014} may provide important pieces of information to verify or exclude certain model inputs. 

All the calculated cross sections and reaction rates are available at \url{http://ocl.uio.no/xfiles}.

\acknowledgments

This work was financed through ERC-STG-2014 under grant agreement no. 637686. Enlightening discussions with G.M. Tveten and Artemis Spyrou are highly appreciated.

\vfill
\end{document}